\documentstyle[12pt,epsfig]{article}

\def\be{\begin{equation}}
\def\ee{\end{equation}}
\def\la{\langle}
\def\ra{\rangle}
\def\IP{\hbox{\rm I\kern -1.6pt{\rm P}}}
\def\IC{{\hbox{\rm C\kern-.58em{\raise.53ex\hbox{$\scriptscriptstyle|$}}
    \kern-.55em{\raise.53ex\hbox{$\scriptscriptstyle|$}} }}}
\def\IN{\hbox{I\kern-.2em\hbox{N}}}
\def\IR{\hbox{\rm I\kern-.2em\hbox{\rm R}}}
\def\ZZ{\hbox{{\rm Z}\kern-.3em{\rm Z}}}
\def\IT{\hbox{\rm T\kern-.38em{\raise.415ex\hbox{$\scriptstyle|$}} }}

\newtheorem{theorem}{Theorem}[section]
\newtheorem{lemma}[theorem]{Lemma}

\newtheorem{proposition}[theorem]{Proposition}
\newtheorem{corollary}[theorem]{Corollary}

\begin{document}

\title{Stability of Solutions of Hydrodynamic Equations Describing the Scaling
Limit of a Massive Piston in an Ideal gas}
\author{E. Caglioti$^{1}$, N. Chernov$^{2}$, and
J.~L.~Lebowitz$^{3}$}
\date{\today}
\maketitle

\footnotetext[1]{Dipartimento di Matematica, Universit\`a di Roma
 ``La Sapienza'' P.le A. Moro 2, 00185 Roma, Italy}
\footnotetext[2]{Department of Mathematics, University of Alabama at
Birmingham, Alabama 35294} \footnotetext[3]{Department of Mathematics,
Rutgers University, New Jersey 08854}

\begin{abstract}
We analyze the stability of stationary solutions of a singular Vlasov
type hydrodynamic  equation (HE).  This equation was derived (under
suitable assumptions) as the hydrodynamical scaling limit of the
Hamiltonian evolution of a system consisting of a massive piston
immersed in  an ideal gas of point particles in a box.  We find
explicit criteria for global stability as well as a class of solutions
which are linearly unstable for a dense set of parameter values.  We
present evidence (but no proof) that when the mechanical system has
initial conditions ``close'' to stationary stable solutions of the HE
then it stays close to these solutions for a time which is long
compared to that for which the equations  have been derived.  On the
other hand if the initial state of the particle system is close to an
unstable stationary solutions of the HE the mechanical motion follows
for an extended time a perturbed solution of that equation:
we find such approximate periodic solutions that are linearly stable.

\end{abstract}

\renewcommand{\theequation}{\arabic{section}.\arabic{equation}}

\section{Introduction}
\label{secI} \setcounter{equation}{0}

The time evolution of a system consisting of a piston of mass $M$
moving parallel to the $x$-axis in a cube containing non-interacting
point particles of unit mass has been studied extensively
\cite{CLS1,CLS2,CL,G,GF,GP,H,KBM,LPS,Li,PG}. After some rescaling of
space and time (by the length of the cube) the problem reduces to that
of a one dimensional system with $N_L$ ($N_R$) particles in the
interval $[0,X]$ (resp., $[X,1])$ where $X(t)$ is the position of the
piston.  The left (right) particles move freely between collisions with
the wall at $x=0$ ($x=1$) and the piston at $x = X(t)$.  At the walls
the velocities of the particles get reversed while at $x = X(t)$ the
outgoing velocity $v^\prime$ is related to the incoming velocity $v$ by
the rules of elastic collisions \cite{CLS2,CL,LPS},
\be
    v^\prime = -\frac{M-1}{M+1}\, v + \frac{2M}{M+1}\, V
      \label{vprime}
\ee
where $V$ is the incoming velocity of the piston.  It follows
from (\ref{vprime}) that $N_L, N_R$, as well as the total kinetic
energy, ${1 \over 2} \sum^N_{i=1} v_i^2 + {1 \over 2} MV^2$ ($N
= N_L+N_R$) are conserved quantities. The dynamics
 of the
system can be reduced to a billiard in a
 $(2N+1)$-dimensional domain
(polyhedron), cf.\ \cite{CL}.  It was shown in
\cite{LPS,CLS1,CLS2}, under certain quite restrictive
conditions on the initial distribution of gas particles, that, in
the limit $N \to \infty$, $M \sim N^{2/3}$, the dynamics of the piston and the gas
satisfy a closed system of Euler-type hydrodynamical equations (HE)
for a time interval $(0,\tau)$ in which any particle had at most two collisions with the piston.

The origin of the scaling $M \sim N^{2/3}$ is as follows. For $N$
particles with velocities of $O(1)$ distributed with density of $O(1)$
in a parallelopiped of length $L$ and crossectional area $A$ the number
of particles colliding with the piston per unit (unrescaled) time and
hence the pressure (from each side) is proportional to $A$.  To ensure
that, on this time scale, the acceleration of the piston stays of
$O(1)$  as $L$, $A$ and $N \sim O(AL)$ grows to infinity it is
necessary to make the mass of the piston grow as $A$.  For a cube this
corresponds to $M \sim N^{2/3}$.  In the rescaled units the number of
collisions experienced by the piston per unit time is of $O(N)$
independent of $A$ and the HE hold for general $M$ as long as $M \sim
N^\alpha$, $\alpha\in (0,1)$, i.e.\ when the kinetic energy of the
piston becomes negligible compared to that of the gas. The time
interval (in the scaled units) during which the derivation of the HE
remains valid depends on $\alpha$ (getting larger as $\alpha \to 1$),
see Remarks 3 and 4 in Sect.~4 of \cite{CLS2}. For $\alpha =2/3$ this
time interval is such that the piston suffers no more than two
collisions with any gas particle. It is however not clear from the
derivation to what extent those equations may actually approximate the
real evolution of the system with large $N$, for longer times.  This
led us to carry out extensive computer simulations of particle systems,
with $M\sim N^{2/3}$, $N$ as large as $27 \times 10^6$ \cite{CL}, and
initial conditions for which the hydrodynamic equations have trivial
solutions $X(t) = 0.5$ and $V(t) = 0$ for all $t>0$.  We found
nevertheless that for certain initial velocity distributions the
trajectory of the piston diverged from these values. In particular it
was observed in these simulations that the particle plus piston system,
after experiencing large random fluctuations, quickly converges to a
more stable regime, in which the piston and the gas undergo regular
slowly damped oscillations lasting a long time. The parameters of those
oscillations (the period, the amplitude, and the rate of damping) seem
to depend little on many details of the initial distributions.

A possible explanation for this behavior is that the mechanical
trajectory of the system with $N$ large but finite, being subjected to
the intrinsic random fluctuations associated with the discrete nature
of the gas particles, behaves as a perturbed solution of the HE, i.e.\
as if the initial conditions were subjected to a small random
perturbation.  There large fluctuations in the particle dynamics
observed experimentally in \cite{CL} may thus be due to the instability
of the HE for the initial conditions considered.  This would imply
that, on the contrary, when the solution of the hydrodynamical
equations is stable, then the mechanical evolution of the system should
remain close to that solution for a very long time.

Here we present evidence, but no proof, for this conjecture. We first
find a class of initial conditions for which the solution of the HE
are stable and note that simulations with such distributions in
\cite{CL} indeed yielded a piston
trajectory, which remained close to the
solution of the HE with $V(t) \sim 0$ .  Furthermore, we observe that the
mechanical  piston trajectories, which do not stay close to the
solution of the HE with the
prescribed initial conditions, appear to follow some perturbed
oscillating solutions of
the HE equations.  These
solutions  appear to act as an ``attractor''
for the HE \footnote{Due to an
obvious time-reversibility of the hydrodynamical equations, there can
be no attractors, strictly speaking, but saddle-type periodic solutions
can very well exist.}. We cannot rigorously prove the existence of
attracting periodic solutions for the HE yet, but we present some approximate
constructions of such solutions.

The paper is organized as follows. In Section 2 we state the
hydrodynamical equations. In Section 3 we prove global stability for a
 class of stationary solution. In Section 4, we use a
perturbative analysis to find sufficient conditions for linear
instability. In Section 5 we investigate a particular family of
stationary solutions which include those used in the \cite{CL}
simulations and show that it contains both linearly stable and unstable
ones, alternating in a very intricate manner. In Section 6, an
approximate construction of periodic solutions of the HE is presented.

\section{Hydrodynamical equations}
\label{secHE} \setcounter{equation}{0}

%The motion of a massive piston in a cylinder filled with an ideal
%gas can be described by a closed system of equations constructed
%in \cite{LPS,CLS1,CLS2}.

Let $X(t)\in (0,1)$ be the position of
the piston at time $t$ and $V(t)$ its velocity.  We
denote the continuum density of the gas in $[0,1] \times \IR$ by a
function $p(x,v,t)$. The HE describing the
time evolution of this, continuum fluid plus piston system, are as
follows.

\begin{itemize}
\item[(H1)] {\em Free motion}. Inside the container the density
satisfies the standard continuity equation for a noninteracting
particle system without external forces:
\be
     \left ( \frac{\partial}{\partial t}+
      v\, \frac{\partial}{\partial x} \right )
        \, p(x,v,t)=0
          \label{pdinside}
\ee
for all $x \in (0,1)$ except $x=0$, $x=1$ and $x=X(t)$.
\end{itemize}

\noindent Equation (\ref{pdinside}) has a simple solution
\be
      p(x,v,t)=p(x-vs,v,t-s)
         \label{pinside}
\ee
for $0<s<t$ such that $x-vr\notin\{0,X(t-r),1\}$ for all $r\in
(0,s)$. Equation (\ref{pinside}) has one advantage over
(\ref{pdinside}): it applies to all points $(x,v)$, including
those where the function $p$ is not differentiable, or even continuous.

\begin{itemize}

\item[(H2)] {\em Collisions with the walls}. At the walls $x=0$
and $x=1$ we have
\be
       p(0,v,t) = p(0,-v,t)
          \label{pwall0}
\ee
\be
       p(1,v,t) = p(1,-v,t)
          \label{pwall1}
\ee

\item[(H3)] {\em Collisions with the piston}. At the piston
$x=X(t)$ we have (this is obtained from (\ref{vprime}) when $M \to \infty$)
\begin{eqnarray}
      p(X(t)-0,v,t) &=& p(X(t)-0,2V(t)-v,t)
      \ \ \ \ \ {\rm for}\ \ v<V(t)\nonumber\\
      p(X(t)+0,v,t) &=& p(X(t)+0,2V(t)-v,t)
      \ \ \ \ \ {\rm for}\ \ v>V(t)
        \label{ponpiston}
\end{eqnarray}
where $v$ represents the velocity after the collision and $2V(t)-v$
that before the collision; and
\be
     X(t) = X(0) + \int_0^t V(s)\, ds
          \label{V=X'}
\ee
is the (deterministic) position of the piston.

\end{itemize}

It remains to describe the evolution of $V(t)$ (which we take to be
left continuous). Suppose the
piston's position at time $t$ is $X$ and its velocity $V$. The
piston is affected by the fluid at $(x,v)$ exerting pressure on it
from
the right ($x=X+0$ and $v<V$) and from the left ($x=X-0$ and $v>V$). Accordingly, we define the density of the
fluid in contact  with the piston (``density on the piston'') by
\be
    q(v,t;X,V)=\left\{\begin{array}{ll}
       p(X+0,v,t)  &  {\rm if}\ \ v<V\\
       p(X-0,v,t)  &  {\rm if}\ \ v>V\\
          \end{array}\right .
            \label{qp}
\ee

\begin{itemize}
\item[(H4)] {\em Piston's velocity}. The velocity $V=V(t)$ of the
piston satisfies the equation
\be
       \int_{-\infty}^{\infty} (v-V)^2\,{\rm sgn}
       (v-V)\, q(v,t;X,V)\, dv =0
            \label{quadraticint}
\ee
\end{itemize}
The origin of eqs. (H1)-(H3) in the particle system is clear.  Eq. (H4)
is essentially a force balance equation---since the rate of collision
of the piston with particles on either side and consequent force on
piston is much larger than mass of the piston when $N/M \to \infty$,
$V$ adjusts instantaneously to make the forces from the two sides
balance exactly. The system of (hydrodynamical) equations (H1)--(H4) is
now closed and, given initial conditions, satisfying (2.3)-(2.8) at
$t=0$, completely determine the functions $X(t)$, $V(t)$ and $p(x,v,t)$
for $t>0$. When the initial conditions do not satisfy these equations
one has to imagine that they become satisfied instantaneously for
$t=0+$.  The existence and uniqueness of solutions of (H1)--(H4) were
proven, under general conditions, in \cite{CLS1,CLS2}. We need only to
assume that the $p(x,v,0)$ is bounded, piecewise differentiable, and
either has a compact support in the $x,v$ plane or decays fast enough
as $|v|\to\infty$.  We also require that $\int p(x,v,0)dv > 0$ for all
$x$.

The HE, like the Vlasov equations for
plasmas, are time-reversible, see \cite{P} and \cite{MP}. They preserve the classical
integrals of motion. The mass of the fluid to the left and to the
right of the piston as well as the total kinetic energy of the fluid
remain constant along any solution.

\begin{itemize}
\item[(D1)]{\em Mass conservation}
$$
   {\cal M}_L=\int_0^{X(t)}\!\int p(x,v,t)\, dv\, dx,
   \ \ \ \ \ \
   {\cal M}_R=\int_{X(t)}^1\!\int p(x,v,t)\, dv\, dx
$$
\item[(D2)]
{\em Energy conservation}
\be
    2E=\int\!\int v^2 p(x,v,t)\,dv\, dx
          \label{tildeFpiston}
\ee
\end{itemize}
Equation (\ref{quadraticint}) also preserves the total momentum
$\int\!\int v\, p(x,v,t)\, dv\, dx$, but the latter changes due to
reflections at the walls. We note that the piston itself does not
contribute to the total momentum and energy of the system in this
model, because its mass and energy vanish, when divided by $N$, in the
limit $N\to\infty$. (The mass, energy and momentum of the fluid all
correspond to the original quantities in the particle system divided by
$N$).

The HE define a dynamics on the domain $ G:=\{(x,v):\ 0\leq x\leq 1\}$
in which every point $(x,v)\in G$ moves freely with constant velocity
and collides elastically with the walls and the piston. Denote by
$(x_{t},v_{t})$ the position and velocity of an arbitrary point at time
$t\geq 0$. Then (H1) translates into $\dot{x}_{t} =v_{t}$ and
$\dot{v}_{t}=0$ whenever $x_{t} \notin\{0,1,X(t)\}$, (H2) becomes
$(x_{t+0},v_{t+0})= (x_{t-0},-v_{t-0})$ whenever $x_{t-0} \in\{0,1\}$,
and (H3) gives
$$
       (x_{t+0},v_{t+0})=(x_{t-0},2V(t)-v_{t-0})
$$
whenever $x_{t-0}=X(t)$. Note that the point $(x_t,v_t)$ moves in $G$
and reflects at the walls and the piston as if those had infinite mass.

The motion of points in $G$ is described by a one-parameter family of
transformations $F^{t}:\, G\to G$ defined by
$F^{t}(x_0,v_0)=(x_{t},v_{t})$ for $t>0$. We will also write
$F^{-t}(x_t,v_t)=(x_0,v_0)$. According to (H1)--(H3), the density
$p(x,v,t)$ satisfies a simple equation
$$
   p(x_{t},v_{t},t)= p(F^{-t}(x_{t},v_{t}),0)= p(x_0,v_0,0)
$$
for all $t\geq 0$. It is easy to see that for each $t>0$ the map
$F^{t}$ is one-to-one and preserves area, i.e. ${\rm det}\, |DF^{t}
(x,v)| =1$. Hence, the family $F^t$ describes an incompressible flow on
$G$ and consequently:
\begin{itemize}
\item[(D3)] {\em Incompressibility}. For any $a<b$ the Lebesgue
measures (areas) of the sets
$$
     \{(x,v):\ a<p(x,v,t)<b,\ 0<x<X(t)\}
$$
and
$$
     \{(x,v):\ a<p(x,v,t)<b,\ X(t)<x<1\}
$$
remain constant in time

\end{itemize}

A particular case in which it is possible to solve equations
(H1)--(H4) analytically, is when the initial
distribution is stationary.  This happens when
$p(x,v,0)$ satisfies two
conditions:

\begin{itemize}
\item[(S1)] {\em Uniformity and symmetry}. The initial density
$p(x,v,0) = p(x,v)$ is of the form
$$
   p(x,v)=\left\{\begin{array}{ll}
   p_L(|v|) & {\rm for}\ \ x<X_0 \\
   p_R(|v|) & {\rm for}\ \ x>X_0
   \end{array} \right .
$$
for all $v$ and $X(0) = X_0$.

\item[(S2)] {\em Pressure balance}. The pressure on the piston from
both sides is equal:
\be
    P_L:=2\int_0^{\infty}v^2p_L(v)\,dv = P_R:=2\int_0^{\infty}v^2p_R(v)\,dv
       \label{PLPR}
\ee
\end{itemize}

Under conditions (S1)--(S2) the equations (H1)--(H4) have a simple
solution: the system remains frozen in its initial state:
\be
   X(t)\equiv X_0,\ \ \ \
   V(t)\equiv 0,\ \ \ \
   p(x,v,t)\equiv p(x,v,0)
     \label{statsol}
\ee
for all $t>0$.

We will analyze in the next three sections the stability of this stationary
solution.  Note that there is no
requirement on the form of  $p_L(|v|)$ or $p_R(|v|)$; all that
 is required is a
balance of forces (2.10).

\section{Globally stable solutions}
\label{secSS} \setcounter{equation}{0}

Here we consider stationary solutions  $p(x,v)$ satisfying (S1)--(S2)
and
an additional monotonicity requirement:
\be
   p_L(\vert v_1\vert)\geq p_L(\vert v_2\vert )\ \ \ \ {\rm and}\ \ \ \
    p_R(|v_1|)\geq p_R(|v_2|)
       \label{monot}
\ee
for all $|v_1|\leq |v_2|$. % For example, the uniform density
%$$
%    p(v) = \left\{\begin{array}{ll}
%   c>0 & {\rm for}\ \ v\in (-a,a) \\
%   0 & {\rm otherwise}
%   \end{array} \right .
%$$
%satisfy these conditions as does the Maxwellian distribution.
%
We claim that such solutions
 are globally stable.
%%%%%%added by emanuele%%%%%
%%%%%%%%%%%%%%%%%%%%%%%%%%%%%%%%%%%%%%%%%%%%%%%%%%%%%%%%%%%%%%%%%%%%%%%%%%%%
This criteria is very similar to the stability criteria for the Vlasov
equation described by Penrose \cite{P} and by Marchioro and Pulvirenti
\cite{MP}.

%%%%%%%%%%%%%%%%%%%%%%%%%%%%%%%%%%%%%%%%%%%%%%%%%%%%%%%%%%%%%%%%%%%%%%%%%%%%

 Before we state our result, we introduce
some notation. Denote by $\|\cdot\|$ the following special norm on
the space of functions on $G$
%changed by emanuele
%the norm was not really introduced in that paper
%
%%%%%%%%%%%%%%%%%%%%%%%%%%%
\be
      \|f(x,v)-g(x,v)\| = \int\!\int |f(x,v)-g(x,v)|\, (1+v^2)\, dv\, dx
        \label{norm}
\ee

\begin{theorem}
Let $p(x,v)$ satisfy {\rm (S1), (S2)} and
{\rm (\ref{monot})}. Then for any $\varepsilon>0$ there exists
$\delta>0$ such that if the initial density $p(x,v,0)$
 satisfies  $\|p(x,v,0)-p(x,v)\|<\delta$, and $X(0)=X_0$ then
\begin{itemize}
\item[{\rm (i)}] $\|p(x,v,t)-p(x,v)\|<\varepsilon$; \item[{\rm
(ii)}] $|X(t)-X(0)|<\varepsilon$
\end{itemize}
for all $t>0$.
\label{tmstab1}
\end{theorem}

%%%%%%%changed by Emanuele
\noindent{\em Proof}. The proof of claim (i) is  %essentially ?$
identical
to the proof of the stability theorem of Marchioro and Pulvirenti
\cite{MP}. We note that their theorem is stated in
the $L_1$ norm, but, in fact it was proven in the  (\ref{norm})
norm. (There is  a minor error  in \cite{MP},
the conditions that assure that the $L_1$ norm and the (\ref{norm}) norm are
equivalent are not explicitely
assumed.)

 %We note that their theorem is stated, mistakenly, in
%the $L_1$ norm, but in fact it was proven in the (\ref{norm})
%norm.
%
%We now prove claim (ii). Our argument only uses two properties of
%the dynamics:

%\begin{itemize}
%\item[(D2)] {\em Energy conservation}. The total energy $E(t)$
%defined by
%$$
%     2 E(t)=\int_{-\infty}^{\infty}\int_0^{X(t)}\
%      v^2p_L(x,v,t)\, dx\, dv+
%      \int_{-\infty}^{\infty}\int_{X(t)}^1\
%      v^2p_R(x,v,t)\, dx\, dv
%\label{3.3}
%$$
%is constant in time.
%\end{itemize}
%

It is clear from (\ref{tildeFpiston}) that, given the position of the
piston $X$ and values of the areas of the level sets, defined in
$(D3)$, the minimal possible value of the total energy for any
phase-space density $\pi(x,v)$, is attained when $\pi(x,v)$ is uniform
in $x$ and monotonically decreasing in $|v|$ in each compartment.

Consider first the case where $\pi(x,v)$ has the same area of the
level sets as some $p(x,v)$ satisfying $(S1)$ and $(S2)$. Then the minimum of the
energy when the piston position is $X$ is attained when
$$
      \pi(x,v)=p_L(vX/X_0)
, \quad 0 < x < X
$$
and
$$
      \pi(x,v)=p_R(v(1-X)/(1-X_0))
, \quad X < x < 1
$$
The minimal total energy is then
\begin{eqnarray*}
  2  E_{\min}(t) &=& \int_{0}^{\infty}\int_0^{X}\
      v^2\pi(x,v)\, dx\, dv+
      \int_{0}^{\infty}\int_{X}^1\
      v^2\pi(x,v)\, dx\, dv \\
      &=& \frac{X_0^3}{X^2}\int_0^{\infty}v^2p_L(v)\, dv+
      \frac{(1-X_0)^3}{(1-X)^2}\int_0^{\infty}v^2p_R(v)\, dv
\end{eqnarray*}
(we used a change of variable $u=vX/X_0$ in the first integral
and $u=v(1-X)/(1-X_0)$ in the second one). Using the pressure
balance (\ref{PLPR}) and denoting $P=P_L=P_R$ gives
$$
   E_{\min}(t)=\frac P2 \, \left(\frac{X_0^3}{X^2}
   +\frac{(1-X_0)^3}{(1-X)^2}\right )
$$
Consider now the above expression as a function of $X$. Its minimum is
attained at the point where $dE_{\min}/dX=0$, i.e.
$$
        \frac{X_0^3}{X^3}=\frac{(1-X_0)^3}{(1-X)^3}
$$
 which is only possible if $X=X_0$. Therefore, the state
 $X=X_0$
provides a unique minimum of the total energy function under
 the
incompressibility constraint (D3). This proves the theorem when the
$p(x,v,0)$ has exactly the same area of the level sets as
$p(x,v)$. For perturbed initial densities $p(x,v,0)$ the above
estimates only hold approximately, hence small changes in the system
are possible, but large changes are not. This proves the
theorem.
\bigskip
Slightly extending the above theorem, let
us assume that the initial density $p(x,v,0)$ satisfies (S1),
(\ref{monot}), but not (S2). If the pressures $P_L$ and $P_R$ in
(\ref{PLPR}) differ by a small amount $\Delta=|P_L-P_R|$, then one
can estimate how far the piston can swing. The piston can move as
long as
$$
        \frac{X_0^3}{X(t)^2}\int_0^{\infty}v^2p(X(t)-0,v,t)\, dv+
      \frac{(1-X_0)^3}{(1-X(t))^2}\int_0^{\infty}v^2p(X(t)+0,v,t)\, dv
      \leq 2E(0)
$$
where $X_0=X(0)$, as before, and $E(0)$ is the initial total
energy:
$$
      2E(0)=  X_0\int_0^{\infty}v^2p_L(v)\, dv+
       (1-X_0)\int_0^{\infty}v^2p_R(v)\, dv
$$
By simple calculations one obtains, to the leading order of $\Delta$,
the following bound on the piston displacements:
$$
    |X(t)-X_0|\leq
    \frac{2\Delta}{3P_L}\left(\frac{1}{X_0}+
    \frac{1}{1-X_0}\right)^{-1}+O(\Delta^2)
$$

\section{Perturbative analysis}
\label{secPA} \setcounter{equation}{0}

Here we consider initial densities $p(x,v,0)$ satisfying (S2) and
the following stricter version of (S1):

\begin{itemize}
\item[(S3)] {\em Full uniformity and symmetry}. The initial density
$p(x,v,0)$ is uniform in $x$ across the entire cylinder, i.e.\
$p(x,v,0)= p_0(|v|)$ for all $v$ and $0<x<1$.
\end{itemize}

We also assume that the piston is initially at the midpoint $X(0)=0.5$.
Of course, under the conditions (S2)--(S3), the hydrodynamical
equations (H1)--(H4) have a simple stationary solution (\ref{statsol}).
On the other hand, we no longer assume monotonicity (\ref{monot}). We
use perturbative analysis to investigate the stability of the
stationary solution (\ref{statsol}).

{}From now on we denote by $p_0(v)=p_0(|v|)$ an initial density
satisfying (S2)--(S3) and by $p(x,v,0)$ a perturbed initial density,
which we write down as
$$
        p(x,v,0) = p_0(|v|) + \varepsilon p_1(x,v,0)
$$
where $\varepsilon p_1(x,v,0)$ is a small perturbation. We will work to
 first order in $\varepsilon$, i.e.\ ignore terms of order
$o(\varepsilon)$. For $t>0$, we decompose the density $p(x,v,t)$ as
$$
        p(x,v,t) = p_0(|v|) + \varepsilon p_1(x,v,t)
$$
We also set $p_1(x,v,t)=p_L(x,v,t)$ for $x<X(t)$ and
$p_1(x,v,t)=p_R(x,v,t)$ for $x>X(t)$.

According to (\ref{quadraticint}), the velocity $V(t)$ of the piston is
given by
$$
      \int_V^{\infty} (v-V)^2\, [p_0(v)+\varepsilon p_L(X,v,t)]\,dv =
      \int_{-\infty}^V (v-V)^2\, [p_0(v)+\varepsilon p_R(X,v,t)]\,dv
$$
where $X=X(t)$ is the position of the piston. Expanding in
$\varepsilon$ and ignoring terms of order $o(\varepsilon)$ gives
$$
       V(t) = \varepsilon\,\,
       \frac{\int_0^{\infty}v^2p_L(X,v,t)\,dv
        - \int_{-\infty}^0 v^2p_R(X,v,t)\,dv}{4\int_0^{\infty}v\, p_0(v)\, dv}
$$
Note that by integrating by parts we obtain (for piecewise smooth $p_0$)
$$
     2\int_0^{\infty}v\, p_0(v)\, dv
     = - \int_0^{\infty} v^2 p_0'(v)\, dv
$$
We define
$$
                h(v) = -p_0'(v)\ \ \ \ \ \ \ \ {\rm for}\ \ v>0
$$
and  for the sake of completeness, set $h(-v)=h(v)$. Then
\be
       V(t) = \varepsilon\,\,
       \frac{\int_0^{\infty}v^2p_L(X,v,t)\,dv
        - \int_{-\infty}^0 v^2p_R(X,v,t)\,dv}{2\int_0^{\infty}v^2h(v)\, dv}
           \label{Vt}
\ee
The function $p_0(v)$ does not have to be differentiable, we
interpret $-h(v)$ here as the generalized derivative of $p_0(v)$. Denote by $\la\cdot\ra_+$ the
integration $\int_0^{\infty}\cdot\, dv$, and by $\la\cdot\ra_-$ the
integration $\int_{-\infty}^0\cdot\, dv$. Then
$$
       V(t) = \varepsilon\,\,
       \frac{\la v^2p_L(X,v,t)\ra_+
        - \la v^2\, p_R(X,v,t)\ra_-}{2\la v^2h(v)\ra_+}
$$

The density of the gas after interaction with the piston is
given by the formulas (\ref{ponpiston}), which imply:
\begin{eqnarray*}
    p(X-0,-v,t) &=& p(X-0,v+2V,t) \\
     &=& p_0(v+2V) + \varepsilon p_L(X,v,t)\\
     &=& p_0(v) + 2Vp_0'(v) + \varepsilon p_L(X,v,t)\\
     &=& p_0(v) - 2Vh(v) + \varepsilon p_L(X,v,t)
\end{eqnarray*}
Here we assume $v>0$ and ignore terms of order $o(\varepsilon)$. Hence,
the ``reflection rule'' can  be written as
$$
   p_L(X,-v,t)=p_L(X,v,t) - h(v)\, \frac{\la v^2p_L(X,v,t)\ra_+
        - \la v^2\, p_R(X,v,t)\ra_-}{\la v^2h(v)\ra_+}
$$
This expression suggests the introduction of new functions:
$$
     q_{R,L}(x,v,t) =  \frac{p_{L,R}(x,v,t)}{h(v)}
$$
and
$$
         \rho(v) = \frac{v^2 h(v)}{\la v^2h(v)\ra_+}
$$
The above expression for $p_L(X,-v,t)$ can now be written as
\be
   q_L(X,-v,t)=q_L(X,v,t) - \la q_L(X,v,t)\, \rho(v)\ra_+
        + \la q_R(X,v,t)\, \rho(v) \ra_-
       \label{qL}
\ee
Similarly, on the other side of the piston,
\be
   q_R(X,v,t)=q_R(X,-v,t) + \la q_L(X,v,t)\, \rho(v)\ra_+
        - \la q_R(X,v,t)\, \rho(v)\ra_-
         \label{qR}
\ee
One can interpret these ``reflection rules'' as follows: the
functions $q_L$ and $q_R$ ``exchange'' their average values with
respect to the ``density'' $\rho(v)$.

Note that $\rho(v)$ is normalized, so that $\la \rho(v) \ra_+ =
1$, but it is not necessarily positive (or even nonnegative).  On the
other hand when
$\rho(v)\geq 0$, i.e. the unperturbed density
$p_0(|v|)$ is nonincreasing, thus satisfying (\ref{monot}). In this
case the stationary solution (\ref{statsol}) is stable, as we already
know by Theorem~\ref{tmstab1}. Here we recover this result by our
perturbative analysis:

\begin{theorem}
The quantity
$$
    Q = \int\!\int \frac{q^2(x,v,t)\, \rho(v)}{|v|}\, dx\, dv
$$
is constant in time, i.e.\ $dQ/dt=0$. Here $q=q_L$ for $x<X$ and
$q=q_R$ for $x>X$.
\end{theorem}

\noindent{\em Proof}. Clearly, $Q$ cannot change just due to the free
motion of the gas or due to collisions with the walls, so we only need
to worry about collisions with the piston. The gas particles colliding
with the piston during an infinitesimal interval $(t, t+dt)$ lie in the
two triangles on the $xv$ plane: $X-v\, dt<x<X$ for $v>0$ and $X<x<X -
v\, dt$ for $v<0$. The outgoing particles lie in similar symmetric
triangles. Hence, during the interval $(t,t+dt)$, the quantity $Q$
decreases by (up to the factor of $dt$)
$$
      \int_0^{\infty} |v|\, \frac{q_L^2(X,v,t)\, \rho(v)}{|v|}\, dv
      + \int_{-\infty}^0 |v|\, \frac{q_R^2(X,v,t)\, \rho(v)}{|v|}\, dv
$$
$$
      = \la q_L^2(X,v,t)\, \rho(v)\ra_+ + \la q_R^2(X,v,t)\, \rho(v)\ra_-
$$
and it increases by
$$
      \int_{-\infty}^0 |v|\, \frac{q_L^2(X,v,t)\, \rho(v)}{|v|}\, dv
      + \int_0^{\infty} |v|\, \frac{q_R^2(X,v,t)\, \rho(v)}{|v|}\, dv
$$
$$
      = \la q_L^2(X,v,t)\, \rho(v)\ra_- + \la q_R^2(X,v,t)\, \rho(v)\ra_+
$$
After substituting (\ref{qL}) and (\ref{qR}) into the above expressions for
$q_L$ and $q_R$ and some manipulations, all changes in $Q$ cancel out
and so it stays constant. $\Box$ \medskip

When $p_0(|v|)$ is strictly decreasing, hence $\rho(v) > 0$, then
$Q$ is a norm in the space of functions. Thus, the above theorem
implies linear stability.

When $p_0(|v|)$ is decreasing, but not strictly, then $\rho(v)\geq 0$,
but there may be regions where $\rho(v)=0$. They correspond to the
intervals where $p_0'=0$, i.e.\ where $p_0$ is constant. On such
intervals, the reflection rules (\ref{qL})--(\ref{qR}) for the
perturbed density $p_{L,R}$ are trivial:
$$
      p_L(X,-v,t)=p_L(X,v,t)\ \ \ \ \ \ {\rm and}\ \ \ \ \ \ \
       p_R(X,v,t)=p_R(X,-v,t)
$$
In this case $p_L$ and $p_R$ cannot grow either. Therefore, we
obtain linear stability for all nonincreasing $p_0(|v|)$.

Next we turn to unstable solutions. The stationary solution for an
initial density $p_0(|v|)$ satisfying (S2)--(S3) is linearly unstable
if some small
perturbations grow exponentially in time, i.e.\ $\|p_1(x,v,t)\| \sim
\Lambda^t$ for some $p_1(x,v,0)$ and $\Lambda>1$. This is equivalent to
having a positive Lyapunov exponent in the subspace spanned by the
function $p_1$ and its images.
% ### the next five lines are new
To investigate the existence of such perturbations we first simplify
the collision rules (\ref{qL}) and (\ref{qR}). Consider the following
``symmetric'' and ``antisymmetric'' linear combinations of $q_L$ and
$q_R$:
$$
   q_+(x,v,t)=[q_L(x,v,t)+q_R(1-x,-v,t)]/2
$$
and
$$
   q_-(x,v,t)=[q_L(x,v,t)-q_R(1-x,-v,t)]/2
$$
They are defined for $x<1/2$. The collision rules
(\ref{qL})--(\ref{qR}) now take form
\be
     q_+(X,-v,t)=q_+(X,v,t)
     \label{q+}
\ee
and
\be
     q_-(X,-v,t)=q_-(X,v,t) - 2\la q_-(X,v,t)\, \rho(v)\ra_+
     \label{q-}
\ee
Hence $q_+$ is simply a periodic function in $t$, so it cannot grow to
infinity or decrease to zero. In other words, it cannot affect the
stability or instability of the hydrodynamical equations. The latter is
determined by $q_-$ alone. So we will only consider $q_-$ and omit
``$-$'' for brevity. Our collision rule then reduces to a single
equation
\be
     q(X,-v,t)=q(X,v,t) - 2\la q(X,v,t)\, \rho(v)\ra_+
     \label{q}
\ee

Next we demonstrate, by example, that densities $\rho_0(|v|)$ for which
the stationary solution (\ref{statsol}) is  unstable do exist.
\\

\noindent{\bf Example}. Let $p_0$ be a rectangular function defined by
\begin{eqnarray}
    p_0(v) = \left \{
    \begin{array}{ll} 1 & {\rm if}\ \ 0.5<|v|<1 \\
    0 & {\rm otherwise}
    \end{array} \right .
     \label{p12}
\end{eqnarray}
This $p_0(v)$ satisfies (S2)--(S3) but not (\ref{monot}). We will show that the
corresponding stationary solution is linearly unstable.

First, the function $h=-p_0'$ is the sum of two delta functions:
$$
        h(v) = -\delta_{0.5} + \delta_1
$$
(and symmetrically for $v<0$). It is easy to compute $\rho$ directly
$$
        \rho = -\frac 13\, \delta_{0.5} + \frac 43\, \delta_1
$$
Now the reflection rule (\ref{q}) implies
\begin{eqnarray*}
   q(-1) &=& -\frac 53\, q(1)+\frac 23\, q(0.5)\\
   q(-0.5) &=& -\frac 83\, q(1)+\frac 53\, q(0.5)
\end{eqnarray*}
Note that only the values $p(x,\pm 0.5,t)$ and $p(x,\pm 1,t)$ will
evolve in a nontrivial way, as specified above, since $h(v)=0$ for all
$v\notin\{1,0.5,-0.5,-1\}$.

We now construct a linear subspace of functions $q=q_-$ that stays
invariant under the above transformations and in which functions
grow exponentially in time (since the $q_+$ component of the
perturbation is irrelevant, we set it to zero). We can simplify
the construction further by assuming that at time $t=0$
$$
     q(x,\pm 1,0) = u_1, \ \ \ \ \ \
     q(x,0.5,0) = u_2, \ \ \ \ \ \
     q(x,-0.5,0) = u_3
$$
with some constants $u_1,u_2,u_3$ (the choice of indices $1,2,3$ is
rather arbitrary). We note that the functions $p_{L,R}(x,v,0)$ are now
piecewise constant and are completely described by the values
$u_1,u_2,u_3$. The space of such perturbations is three-dimensional.

It is easy to see that at time $t=1$ the functions $p_{L,R}$ will again
be constant on the same intervals, hence they will be described by some
other constants $u_1',u_2',u_3'$. Our collision rule (\ref{q}) implies
that the vectors ${\bf u}'=(u_1',u_2',u_3')^T$ and ${\bf
u}=(u_1,u_2,u_3)^T$ are related by a linear transformation
$$
         {\bf u}'={\bf A}{\bf u}
$$
where $\bf A$ is a $3\times 3$ matrix:
$$
     {\bf A} = \frac 13 \, \left ( \begin{array}{rrr}
     -5 & 2 & 0 \\
     0 & 0 & 3 \\
     -8 & 5 & 0 \\
     \end{array} \right )
$$
After that, the evolution will proceed periodically -- the vector
$\bf u$ will be multiplied by the matrix $\bf A$ at times
$t=1,2,3,\ldots$. The matrix $\bf A$ has three real eigenvalues:
$$
     \lambda_{1,2}=\frac{-4\pm\sqrt{7}}{3}
     \ \ \ \ \ {\rm and}\ \ \ \ \ \
     \lambda_3=1
$$
The largest eigenvalue $\lambda=-(4+\sqrt{7})/3 \approx -2.215$ has the
following (unit) eigenvector:
$$
    {\bf u}=(0.4472,\, -0.3680,\, 0.8152)
$$
This eigenvector spans a one-dimensional subspace in the space of
perturbation densities, which is invariant over the time period $t=1$
and in which the corresponding perturbations are expanded by a factor
$|\lambda|\approx 2.215$. Roughly, the perturbations double over one
period.

To explore the above periodic growth of perturbations, we note that the
piston velocity is given by
\begin{eqnarray*}
     V(t) & = & \frac{\varepsilon}{2}\, \Big (
      \la q_L(X,v,t)\rho(v)\ra_+-\la q_R(X,v,t)\rho(v)\ra_- \Big ) \\
      & = & \varepsilon \la q(X,v,t) \ra_+
\end{eqnarray*}
Hence in our example, during the time interval $0<t<1$
$$
      V= \frac{\varepsilon}{3}\, \left (
      4u_1-u_3 \right ) = 0.9736\,\varepsilon
$$
During the next time interval $1<t<2$ we have
$$
      V= \frac{\varepsilon}{3}\, \left (
      4u_1'-u_3' \right ) = -2.156\, \varepsilon
$$
and so on. Hence, over a unit period of time, the piston velocity grows
by a factor of $|\lambda|=2.215$ and changes sign -- the piston starts
its movements back and forth (oscillations) that increase exponentially
in time.

We note that the same density (\ref{p12}) was studied in \cite{CL}
where the trajectory of the piston was computed after an initial
configuration of gas molecules was selected randomly from the
distribution $p_0(v)$ given in (\ref{p12}). It was found
\cite{CL} that the piston indeed made oscillations which increased
exponentially in time. The piston's velocity grew as const$\cdot R^t$
with $R\approx 1.6$. This experimental estimate is in a reasonable
agreement with our calculation of the largest eigenvalue $\approx
2.215$.

Next, we modify the unstable perturbations $q$ found above and
make them smooth (rather than piecewise constant) functions of
$v$.

We will be looking for the function $q$ of the form
$$
         q(x,v,t)=C(v)\, e^{z(t-x/v)}
$$
where $z$ is a complex constant. Note that due to (\ref{pinside})
the function $q$ (with $v$ fixed) can only depend on $t-x/v$. We
chose the exponential form in order to investigate the existence
of solutions of the linear equation which grow exponentially with
time. Also, for convenience, we introduce the new space coordinate
$y$ in the following way: for all $v>0$ and $x<0.5$ we set
$y=x+0.5$, for $v<0$ and $x<0.5$ we set $y=0.5-x$, for $v>0$ and
$x>0.5$ we set $y=x-0.5$, and for $v<0$ and $x>0.5$ we set
$y=1.5-x$. The so designed coordinate $y$ assumes value zero when
a point $(x,v)\in G$ moving under $F^t$ reflects off the piston,
then grows to 0.5 when the point travels to the wall, and grows
further from 0.5 to 1 when the point travels from the wall back to
the piston.

In the new coordinate $y$, we will be looking for the function $q$ of
the form
$$
         q(y,v,t)=C(|v|)\, e^{z(t-y/|v|)}
$$
More precisely, let
$$
       q(y,\pm 1,t)=C(1)\, e^{z(t-y)}
$$
$$
       q(y,\pm 0.5,t)=C(0.5)\, e^{z(t-2y)}
$$
Recall that $p_0(|v|)$ is the characteristic function of the interval
$[0.5,1]$.

Now, the reflection rule (\ref{q}) implies
$$
    C(1) = -\frac 53\, C(1)\, e^{-z} +\frac 23\, C(0.5)\, e^{-2z}
$$
$$
    C(0.5) = -\frac 83\, C(1)\, e^{-z} +\frac 53\, C(0.5)\, e^{-2z}
$$
We need to find $z$ for which the above system of equations has a
nontrivial solution. Put $e^z=\lambda$ and introduce an auxiliary
variable $D(0.5)=C(0.5)\, e^{-z}$. Now the above system can be
rewritten as
\begin{eqnarray*}
       \lambda C(1) &=& -\frac 53\, C(1) +\frac 23\, D(0.5)\\
       \lambda D(0.5) &=& C(0.5)\\
       \lambda C(0.5) &=& -\frac 83\, C(1) +\frac 53\, D(0.5)
\end{eqnarray*}
Hence, $\lambda$ is an eigenvalue of the matrix of coefficients
$$
     \frac 13 \, \left ( \begin{array}{rrr}
     -5 & 2 & 0 \\
     0 & 0 & 3 \\
     -8 & 5 & 0 \\
     \end{array} \right )
$$
which is the same matrix $\bf A$ that we encountered before. We
take its leading eigenvalue $|\lambda| > 1$ and set
$$
         z = \ln |\lambda | +i\pi
$$
The function $q$ now takes form
$$
      q(y,v,t) = \pm C(|v|)\, |\lambda|^{t-y/|v|}\cos \pi(t-y/|v|)
$$
where $C(0.5)$ and $C(1)$ are the coordinates of the leading
eigenvector, and we only take the real part, for obvious reasons.
Since $|\lambda|>1$, we have an exponential growth of
perturbations and thus linear instability. This gives us smooth unstable
perturbations.

We now generalize the above construction to arbitrary nonmonotonic
initial densities $p_0$. Let $p_0(v)$ satisfy (S2)--(S3) but not
(\ref{monot}). We will be looking for perturbations of the form
\be
         q(y,v,t) = C(|v|)\, e^{z(t-y/|v|)}
           \label{qCexp}
\ee
with the same convention on $y$ as before. The reflection rule
(\ref{q}) leads to (cancelling $e^{zt}$)
$$
         C(v) = C(v)\, e^{-z/v} -
     2\int_0^{\infty}C(v)\, e^{-z/v}\rho(v)\, dv
$$
for all $v>0$. Denoting
$$
      D=-2\int C(v)\, e^{-z/v}\rho(v)\, dv
$$
gives immediately
$$
      C(v) = \frac{D}{1-e^{-z/v}}
$$
Thus, we not only eliminated $t$ but determined the function $C(v)$ up
to a constant factor. The above solution exists if
$$
        D=-2\int_0^{\infty}\frac{D\, e^{-z/v}\rho(v)}{1-e^{-z/v}}\, dv
$$
or, cancelling $D$,
\be
        \int_0^{\infty}\frac{\rho(v)}{1-e^{z/v}}\, dv = \frac 12
       \label{zexp}
\ee
If this equation has a solution $z$ with Re$(z)>0$,  we immediately
obtain an unstable perturbation (\ref{qCexp}). Otherwise our
construction of unstable perturbations does not work.

Unfortunately, it does not seem to be easy to solve equation
(\ref{zexp}) for particular functions $\rho(v)$ or even to determine if
it has solutions with a positive real part, as we will demonstrate in
the next section.

Next we mention an important property of (\ref{zexp}). Let us denote
\be
    F(z):=\int_0^{\infty}\frac{\rho(v)}{1-e^{z/v}}\, dv - \frac 12
      \label{F(z)}
\ee

\begin{lemma} $F(z)+F(-z)=0$ for all $z\in\IC$.
\label{lmsymm}
\end{lemma}

\noindent{\em Proof}.
\begin{eqnarray*}
  F(z)+F(-z) & = & \int_0^{\infty}\frac{\rho(v)}{1-e^{z/v}}\, dv
 - \int_0^{\infty}\frac{e^{z/v}\rho(v)}{1-e^{z/v}}\, dv -1 \\
  & = & \int_0^{\infty}\frac{(1-e^{z/v})\rho(v)}{1-e^{z/v}}\, dv - 1 \\
  & = & 0 \ \ \ \ \ \ \ \ \ \ \ \ \ \ \ \ \ \ \ \ \ \ \ \ \ \ \ \ \ \
  \ \ \ \ \ \ \ \ \ \ \ \ \ \Box
\end{eqnarray*}

As a result, the existence of a solution of (\ref{zexp}) with Re$(z)>0$
is equivalent to that of a solution with Re$(z)<0$. The alternative is
when all the solutions lie on the imaginary axis Re$(z)=0$.

\section{A special family of densities}
\label{secSFD} \setcounter{equation}{0}

Here we investigate a family of rectangular densities
\begin{eqnarray}
    p_0(v) = \left \{
    \begin{array}{ll} 1 & {\rm if}\ \ r<|v|<1 \\
    0 & {\rm otherwise}
    \end{array} \right .
     \label{p1r}
\end{eqnarray}
where $0<r<1$ is the parameter of our family. Note that our example
(\ref{p12}) is a particular case of (\ref{p1r}) with $r=1/2$. It is
easy to compute
$$
      h(v)=-p_0'(v)=\delta_r(v)-\delta_1(v)
$$
and
$$
   \rho(v) = \frac{v^2h(v)}{\int_0^{\infty}v^2h(v)\, dv}
   =\frac{1}{1-r^2} \, \left [
   \delta_1(v) - r^2 \delta_r(v) \right ]
$$
Since $h(v)=0$ for all $v\notin \{1,r, -1,-r\}$, we only consider
perturbations $q(x,v,t)$ defined for $v=1,r,-1,-r$. The reflection rule
(\ref{q}) now gives
\begin{eqnarray*}
   q(X,-1,t) &=& -\alpha\, q(X,1,t)+\beta\, q(X,r,t)\\
   q(X,-r,t) &=& -\gamma\, q(X,1,t)+\alpha\, q(X,r,t)
\end{eqnarray*}
where
\be
   \alpha = \frac{1+r^2}{1-r^2},\ \ \ \ \ \
   \beta = \frac{2r^2}{1-r^2},\ \ \ \ \ \
   \gamma = \frac{2}{1-r^2}
     \label{abg1}
\ee
While we cannot construct unstable perturbations for arbitrary
irrational values of $r$, it is relatively easy to investigate the case
of rational $r$. But even in this case, we can only provide partial
answers leaving some questions open.

\begin{figure}[h]
\centering \epsfig{figure=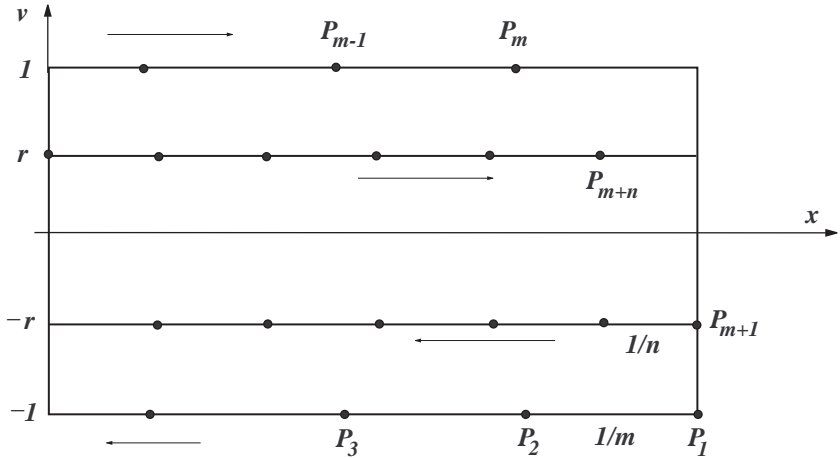}\caption{The construction of points
$P_i$. Here $m=7$ and $n=12$.}
\end{figure}

Let $r=m/n$ be a rational number with $1\leq m <n$. Now (\ref{abg1})
takes form
\be
    \alpha = \frac{n^2+m^2}{n^2-m^2},\ \ \ \ \ \
    \beta = \frac{2m^2}{n^2-m^2},\ \ \ \ \ \
   \gamma = \frac{2n^2}{n^2-m^2}
     \label{abg2}
\ee
To investigate the evolution of perturbations $q(x,v,t)$ as $t$ grows,
we consider $n+m$ points $P_i\in G$, $1\leq i\leq n+m$, shown on
Fig.~1. The points $P_i$ are defined as follows:
$$
   P_i = \left \{ \begin{array}{ll}
   (-1,0.5-(i-1)/m) & {\rm for}\ \ 1\leq i< m/2+1 \\
   (1,(i-1)/m-0.5) & {\rm for}\ \  m/2+1 \leq i \leq m \\
   (-r,0.5-(i-m-1)/n) & {\rm for}\ \ m< i < m+ n/2 +1 \\
   (r,(i-m-1)/n-0.5) & {\rm for}\ \ m+ n/2 +1 \leq i \leq m+n
    \end{array} \right .
$$
It is crucial to observe that the points $P_i$ move under the dynamics
in a periodic fashion. In a time period $\Delta t=1/m$, the point $P_i$
is mapped to $P_{i+1}$ for all $1\leq i<m$ and all $m+1\leq i<m+n$.
Also, $P_m$ moves to the piston, gets reflected off it and lands on
$P_1$. Likewise, $P_{m+n}$ moves to the piston, gets reflected off it
and lands on $P_{m+1}$. Therefore, the time shift $\Delta t$ permutes
the points $P_i$, $1\leq i\leq m+n$, in two independent cycles. The
reason why we combine the two cycles together is that they are linked
by the reflection rule, as we will see shortly.

For each $i$, denote $q_i(t) = q(P_i,t)$. Then we have
$$
    q_i(t+\Delta t) = q_{i-1}(t)
$$
for all $2\leq i\leq m$ and $m+2 \leq i\leq m+n$. The reflection rule
now implies
\begin{eqnarray*}
   q_1(t+\Delta t) &=& -\alpha\, q_{m}(t)+\beta\, q_{m+n}(t)\\
   q_{m+1}(t+\Delta t) &=& -\gamma\, q_{m}(t)+\alpha\, q_{m+n}(t)
\end{eqnarray*}
Thus, the vector ${\bf q}(t) = (q_1(t),\ldots,q_{n+m}(t))$ is updated
at time $t+\Delta t$ by the rule
$$
      {\bf q}(t+\Delta t) = {\bf B} \, {\bf q}(t)
$$
where $\bf B$ is an $(n+m)\times (n+m)$ matrix,
$$
    {\bf B} = \left ( \begin{array} {cccccccc}
     0 & \cdots &  & -\alpha &  & \cdots &  & \beta \\
     1 & \ddots  &  &      &  &        &  &        \\
     \vdots & \ddots & 0 & \vdots & \vdots &        &  & \vdots  \\
       &        & 1 &  0  &  &  \cdots  &  &       \\
     0  &  \cdots &  & -\gamma & 0 &  \cdots &  & \alpha \\
       &         &  &       & 1 &  0     & &        \\
     \vdots &   &  & \vdots &   & \ddots & \ddots & \vdots  \\
     0  & \cdots &  & 0 &   & \cdots & 1 & 0
     \end{array} \right )
$$
We conclude that the existence of unstable perturbations $q(t)$ is
equivalent to the existence of an eigenvalue $\lambda$ of $\bf B$ such
that $|\lambda|>1$. The characteristic polynomial of the matrix $\bf B$
is
\be
    P(\lambda) = \lambda^{m+n} + \alpha \lambda ^n
    - \alpha \lambda^m - 1
\ee
where $\alpha = (n^2+m^2)/(n^2-m^2)$ is defined in (\ref{abg2}).

\medskip
\noindent{\em Remark}. Interestingly, the equation (\ref{zexp}) can be
reduced to $P(\lambda)=0$ as well. Indeed, it is easy to see that
$$
    \int_0^{\infty}\frac{\rho(v)}{1-e^{z/v}}\, dv
    = \frac{1}{1-r^2} \, \left [
    \frac{1}{1-e^z} - \frac{r^2}{1-e^{z/r}}
    \right ]
$$
Now the substitution $\lambda = e^{z/m}$ and some algebraic
manipulations show that Eq.\ (\ref{zexp}) is equivalent to
$P(\lambda)=0$.
\\

It is easy to see that if $\lambda$ is a root of $P(\lambda)$, then so
is $1/\lambda$ (this reciprocability also follows from
Lemma~\ref{lmsymm}). Thus, the existence of unstable perturbations is
equivalent to the existence of eigenvalues of $\bf B$ that do not lie
on the unit circle $|\lambda| =1$.

If an eigenvalue $|\lambda|>1$ of $\bf B$ exists, then the
perturbations in the corresponding eigenspace grow by the factor of
$|\lambda|$ over the time period $\Delta t=1/m$. Hence, the expansion
factor per unit time would be $\Lambda = |\lambda|^m$.

\begin{theorem}
Let $r=m/n$ be a rational number with an even denominator $n$ (hence,
$m$ is odd). Then there is a unique eigenvalue of $\bf B$ such that
$\lambda <-1$. This eigenvalue has multiplicity one. The expansion
factor per unit time $\Lambda_r = |\lambda|^m$ depends on $r$
continuously, and we have, asymptotically,
\be
        \Lambda_r = 1+{\rm const}\cdot r^{3/2}+{\cal O}(r^2)
        \ \ \ \ \ \ \ {\rm as}\ \ r\to 0
           \label{Lambdar0}
\ee
and
\be
       \Lambda_r \sim \frac{\rm const}{1-r}
       \ \ \ \ \ \ \ \ {\rm as}\ \ r\to 1
           \label{Lambdari}
\ee
\label{tmr}
\end{theorem}

\noindent{\em Proof}. One can easily check that, under the conditions
of the theorem, $P(-1)>0$ and $P(-\infty)<0$, hence a root $\lambda<-1$
exists. Next,
$$
   P'(\lambda) = [(n+m)\lambda^n + \alpha n\lambda^{n-m}
   -\alpha m]\, \lambda^{m-1}
$$
and so $P'(-1)<0$ and $P'(-\infty)>0$. Now let $Q(\lambda) =
(n+m)\lambda^n + \alpha n\lambda^{n-m} -\alpha m$, then
$$
   Q'(\lambda) = [(n+m) \lambda^m + \alpha (n-m)]\,
     n\lambda^{n-m-1}
$$
and so clearly $Q'(\lambda)<0$ for all $\lambda<1$. Putting these facts
together proves the uniqueness and the simplicity of the root
$\lambda<0$.

The equation $P(\lambda)=0$ can be rewritten in terms of $\Lambda_r =
|\lambda|^m$ as follows:
\be
   -\Lambda_r^{1+1/r} + \frac{1+r^2}{1-r^2}\Lambda_r^{1/r}
   + \frac{1+r^2}{1-r^2}\Lambda_r - 1 = 0
       \label{Lambdar}
\ee
Now the continuity of $\Lambda_r$, as a function of $r$, is obvious.
Note that our argument is only valid when $r=m/n$ with an even $n$ and
an odd $m$, because this parity condition dictates the signs in
(\ref{Lambdar}).

To prove (\ref{Lambdar0}), one can substitute $\Lambda_r=1+\varepsilon$
into (\ref{Lambdar}) and expand all the terms in Taylor series, the
calculation is then straightforward and we omit it. The proof of
(\ref{Lambdari}) is similar. $\Box$ \medskip

\begin{figure}[h]
\centering \epsfig{figure=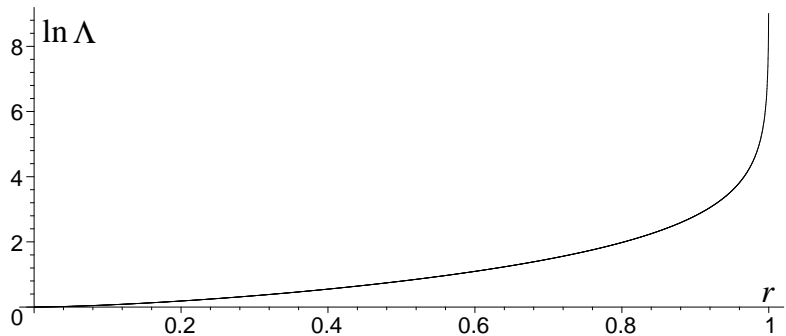}\caption{$\log\Lambda_r$ as a
function of $r$.}
\end{figure}

Figure 2 presents the graph of the Lyapunov exponent $\log\Lambda_r$ as
a function of $r$.

\begin{lemma}
Let $z$ be a solution of (\ref{zexp}) such that $|e^z|\neq 1$ and
$e^z\in\IR$. Then $dF/dz\neq 0$ (in fact, $dF/dz$ is a real negative
number).
\end{lemma}

\noindent{\em Proof}. A direct calculation shows that
$$
    \frac{dF}{dz} = -\frac{(1+r+r^2)(1+e^z)^2 + r^3(1-e^z)^2}
        {4r^2(1+r)(1-e^z)^2}
$$
which proves the lemma. $\Box$ \medskip

For any $r=m/n$ with an even $n$ and odd $m$, the corresponding
solution $e^z=-\Lambda_r$ satisfies the conditions of the above lemma.
Hence, this solution changes continuously with $r$, and so we get the
following

\begin{corollary}
For every $r=m/n$ with an even $n$ and an odd $m$ there is an interval
$(r-\varepsilon,r+\varepsilon)$ in which all parameter values have
unstable perturbations.
\end{corollary}

Therefore, unstable perturbations exist for an open and dense set of
parameter values $0<r<1$. One would naturally wonder if all $r$'s have
unstable perturbations. The answer is, surprisingly, negative:
\medskip

\noindent{\bf Fact 5.4} {\em For the density} (\ref{p1r}) {\em
with $r=1/3$, there are no solutions $z$ of (4.9) with $Re(z)>0$,
hence there are not solutions of the linearized equation which
grow exponentially with time}.
% ### the following sentence has been deleted:
% Thus the corresponding stationary solution is linearly stable.}
\medskip

\noindent{\em Proof}. The characteristic equation
$$
      \lambda^4 + \frac 54 \lambda^3 - \frac 54 \lambda - 1 = 0
$$
has two real roots ($\lambda=\pm 1$) and two complex roots. The complex
roots are, on the one hand, conjugate and, on the other, satisfy the
reciprocability rule: $P(\lambda)=0$ if and only if $P(1/\lambda)=0$.
Hence, they must belong to the unit circle $|\lambda|=1$. $\Box$
\medskip

It is interesting to know if other rational parameter values $r=m/n$
with odd $n$ are also stable. We have examined the values $r=1/n$ for
small odd values of $n=5,7,\ldots,31$ numerically (by using MATLAB) and
always found that all the roots of $P(\lambda)$ belonged to the unit
circle. Therefore, we conjecture that the values $r=1/n$ with odd $n$
are stable.

On the other hand, the values $r=m/n$ with an odd $n$ but $m>1$ appear
to be unstable. For $r=2/3$, $2/5$, $3/5$, $3/7$ we found, again
numerically (by using MATLAB), roots $\lambda$ such that $|\lambda|
>1$. All those roots are complex, for example for $r=2/3$ they are
$\lambda = -0.3778 \pm 1.7173\, i$. It remains to determine
theoretically whether all rational values $r=m/n$ with $m>1$ are
unstable, we leave this question open.

Fact 5.4 seems to disagree with Theorem~\ref{tmr}. Indeed, let $p_0(v)$
be the rectangular density (\ref{p1r}) corresponding to $r=1/3$ and
$p(x,v,0) = p_0(v) + \varepsilon p_1(x,v,0)$ an arbitrary perturbation
with an infinitesimally small $\varepsilon$. According to Fact 5.4,
this perturbation cannot grow exponentially in time. On the other hand,
let us approximate $1/3$ by a rational number $r=m/n$ with an even $n$.
Denote by $p_0^{\ast}(v)$ the corresponding rectangular density
(\ref{p1r}) for the chosen $r=m/n$. Then we have
\be
      p(x,v,0) = p_0^{\ast}(v) + \varepsilon p_2(x,v,0)
      \ \ \ {\rm with}\ \ \ p_2 = p_1 + (p_0 - p_0^{\ast})/\varepsilon
        \label{p1p2}
\ee
Hence, if $|r-1/3| < \varepsilon$, then $(p_0 -
p_0^{\ast})/\varepsilon$ is of order one (in the $L^1$ metric), and
$p(x,v,0)$ becomes an $\varepsilon$-perturbation of the density
$p_0^{\ast}(v)$. As such, it ``must'' grow exponentially in time
according to Theorem~\ref{tmr}. This apparent disagreement requires an
explanation, which we provide next.

We recall that smooth unstable perturbations are given by the general
formula (\ref{qCexp}). For the rectangular density (\ref{p1r}), the
velocity $v$ in this formula only takes two values, $|v|=r$ and
$|v|=1$, hence the factor $C(|v|)$ takes two values, as well, and so
plays little role. For simplicity, we set $|v|=1$ and ignore the
constant factor $C(|v|)=C(1)$. Now the (real part of) unstable
perturbations is described by
\be
           q(y,1,t) =\,{\rm Re}\, e^{z(t-y)}
           = e^{({\rm Re}\, z) (t-y)}
           \cos[({\rm Im}\, z) (t-y)]
              \label{qy1t}
\ee
A similar formula holds for $|v|=r$, and we omit it. Now recall that
for any rational $r=m/n$ we have $e^{z/m} = \lambda$, where $\lambda <
-1$ is the eigenvalue of $\bf B$ described by Theorem~\ref{tmr}.
Therefore, Re$\, z=m\log |\lambda| = \log\Lambda$ and Im$\, z = \pm
m\pi$.

We see that the real part of $z$ changes continuously with $r=m/n$ but
the imaginary part does not. In particular, when $r=m/n$ is close to
$1/3$ and $n$ is even, both $m$ and $n$ have to be large, so that
$m\to\infty$ and $|\,{\rm Im}\, z| \to \infty$ as $r\to 1/3$. In terms
of the perturbation (\ref{qy1t}), the growth of $|\,{\rm Im}\, z|$, as
$r$ approaches $1/3$, implies that the function $q(y,1,t)$ becomes
highly oscillatory, and so does the corresponding initial unstable
perturbation $p(x,v,0)=h(v)q(x,v,0)$. Thus, the linear subspace of
unstable perturbations (along which exponential growth takes place)
becomes nearly orthogonal to any given function, in particular to
$p_2(x,v,0)$ defined in (\ref{p1p2}).

This explains the above ``disagreement''. The density $p_2$ does grow
exponentially in time for any $r=m/n$ with an even $n$, but, as $r\to
1/3$, the projection of $p_2$ onto the unstable subspace corresponding
to the positive Lyapunov exponent $\log\Lambda_r>0$ becomes small and
vanishes in the limit, hence the exponential growth is not visible
during a long initial interval of time. In the limit $r\to 1/3$ that
``initial interval'' becomes infinite and the instability evaporates.

One can also reverse this line of argument. Indeed, when $\varepsilon$
is not infinitesimally small but finite, the representation
(\ref{p1p2}) implies that any perturbation $p(x,v,0)$ of the
rectangular density (\ref{p1r}) for {\em any} $0<r<1$ will eventually
grow exponentially fast in time (because any $r\in (0,1)$ can be
approximated by rational numbers $m/n$ with even $n$). We checked this
conclusion experimentally and found that it was indeed correct.

\vspace*{5mm} \centerline{\epsfbox{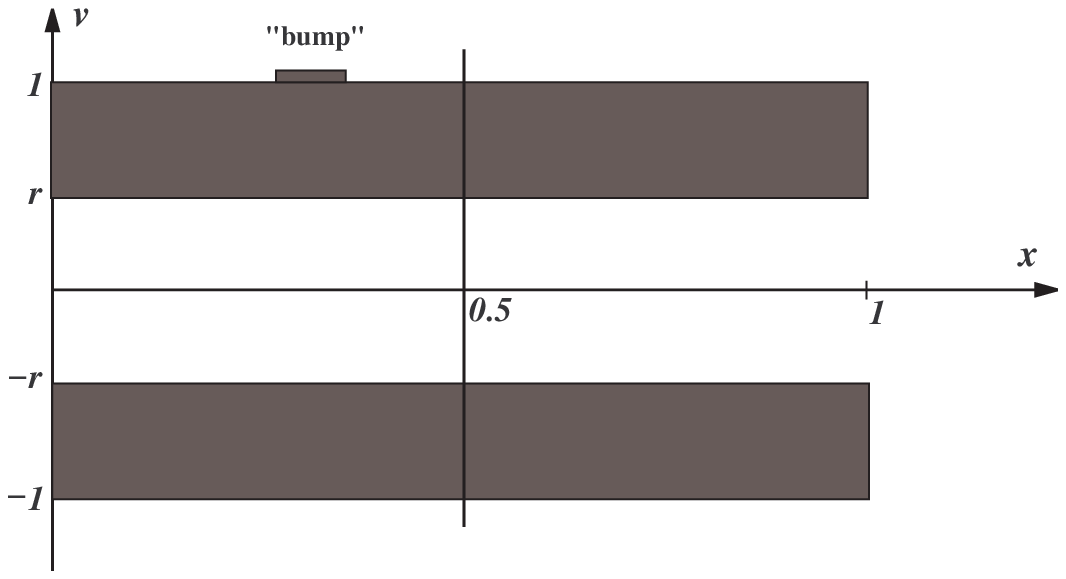}}

\begin{center}
Figure 3: Initial rectangular density (\ref{p1r}) perturbed by a
``bump''.
\end{center}
\vspace*{5mm}

To investigate the instability experimentally, we solved the
hydrodynamical equations (H1)--(H4) numerically starting with a
perturbed rectangular density (\ref{p1r}) shown on Fig.~3. The initial
density $p(x,v,0)$ takes the value one on the black region and zero
elsewhere. The small `bump'' on the top left edge of the upper
rectangle represents the perturbation. The area of the bump in our
experiments was less than 1\% relative to the total area of each black
rectangle.

\vspace*{10mm} \centerline{\epsfbox{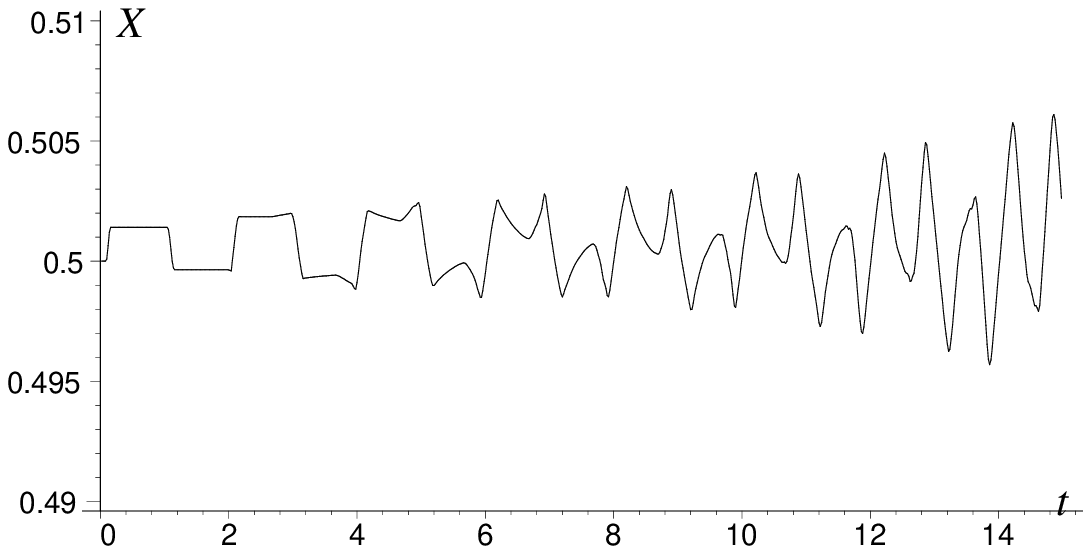}}

\begin{center}
Figure 4: Piston's trajectory for a perturbed rectangular density
with $r=1/3$.
\end{center}

\centerline{\epsfbox{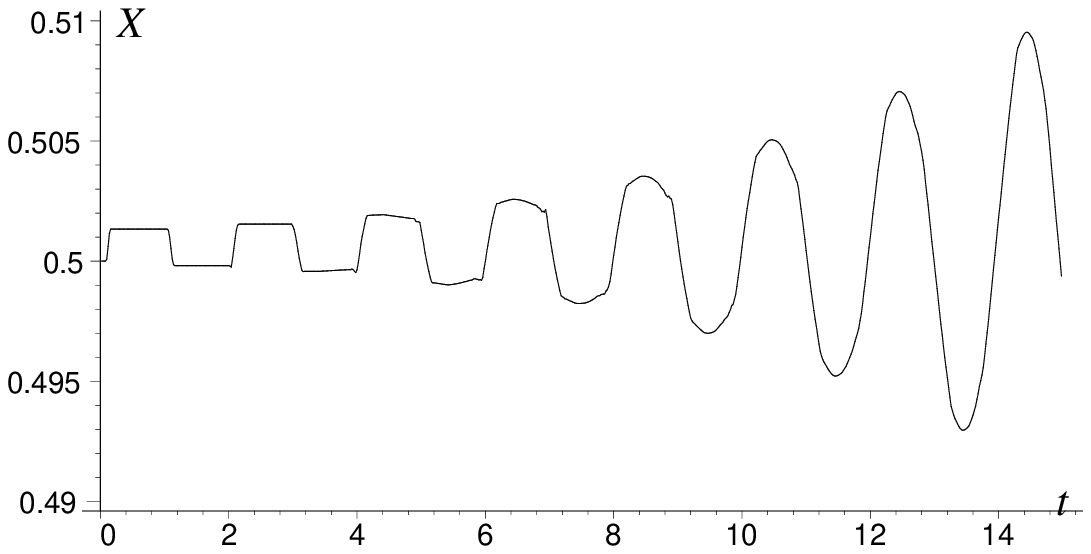}}

\begin{center}
Figure 5: Piston's trajectory for a perturbed rectangular density
with $r=1/4$.
\end{center} \vspace*{5mm}

Figures 4 and 5 show typical trajectories of the piston $X(t)$ for
$r=1/3$ and $r=1/4$, respectively. One can see that by the time $t=15$
the piston's motion shows signs of exponential instability in both
cases, but there is a notable difference. For $r=1/4$ the piston just
swings back and forth with a monotonically increasing amplitude, as
time goes on. For $r=1/3$, the amplitude of oscillations grows slower
but the frequency increases quickly. The higher frequency of
oscillations of $X(t)$ for $r=1/3$ probably reflects the oscillatory
structure of unstable perturbations for rational $r=m/n$ approximating
$1/3$.

\section{Periodic solutions of the hydrodynamical equations}

Here we discuss the long-term behavior of our system in the
unstable regime.

In our previous work \cite{CL} we reported the results of computer
simulations of the piston dynamics in an ideal gas with many (up to 27
million) particles. The initial configuration of particles was selected
randomly with the average density (\ref{p12}), see \cite{CL} for
details. A typical trajectory of the piston $X(t)$ found in our
experiments is shown here on Fig.~6. One can see that during the
initial interval of time $0<t<8$ the piston moves back and forth with
an exponentially increasing amplitude, which is consistent with our
analysis in Section~\ref{secPA}, where the density (\ref{p12}) was
proven to be unstable.

\vspace*{5mm} \centerline{\epsfbox{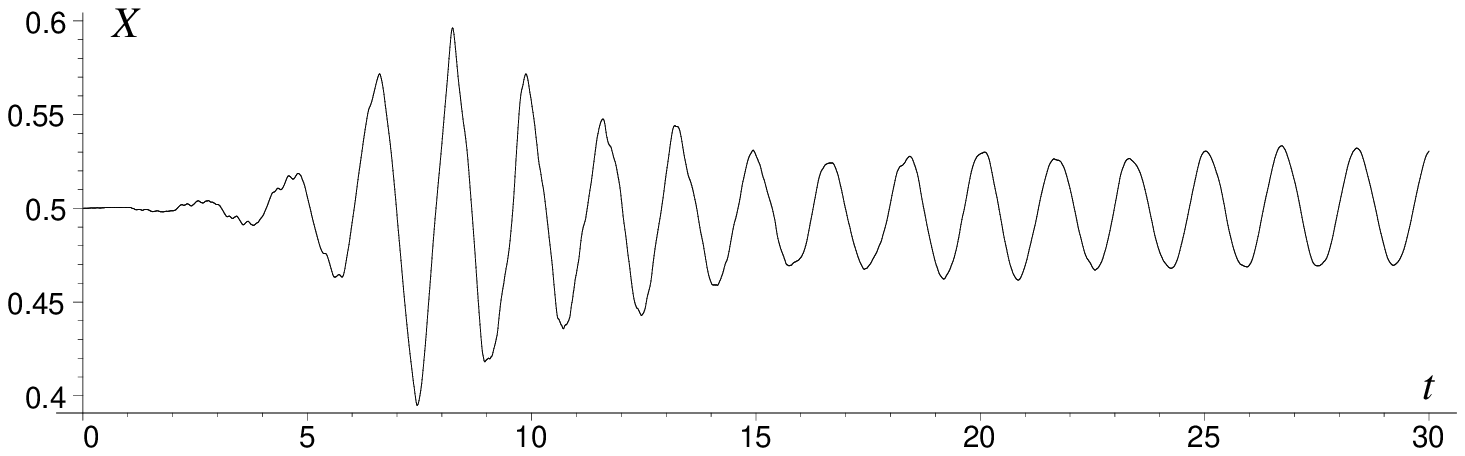}}

\begin{center}
Figure 6: Piston's trajectory in the mechanical model with $10^6$
particles.
\end{center}
\vspace*{5mm}

Later on, however, at times $8<t<15$, the amplitude of the piston's
oscillations decreases to a certain constant value (nearly a half of
its maximum attained at $t=8$). Then the piston's oscillations become
very stable and continue almost unchanged for a very long time, up to
$t=50$ or $100$, with a very slowly decreasing amplitude.

On the other hand, we have solved the hydrodynamical equations
(H1)--(H4) numerically, starting with the same initial density
(\ref{p12}) perturbed by a bump shown on Fig.~3. Figure~7 presents the
resulting trajectory of the piston. One can see that it behaves almost
identically to the simulated trajectory of the piston shown on Fig.~6.
Thus, not only the initial instability, but also the long term behavior
of the simulated piston trajectory match those of perturbed solutions
of the hydrodynamical equations.

\vspace*{5mm} \centerline{\epsfbox{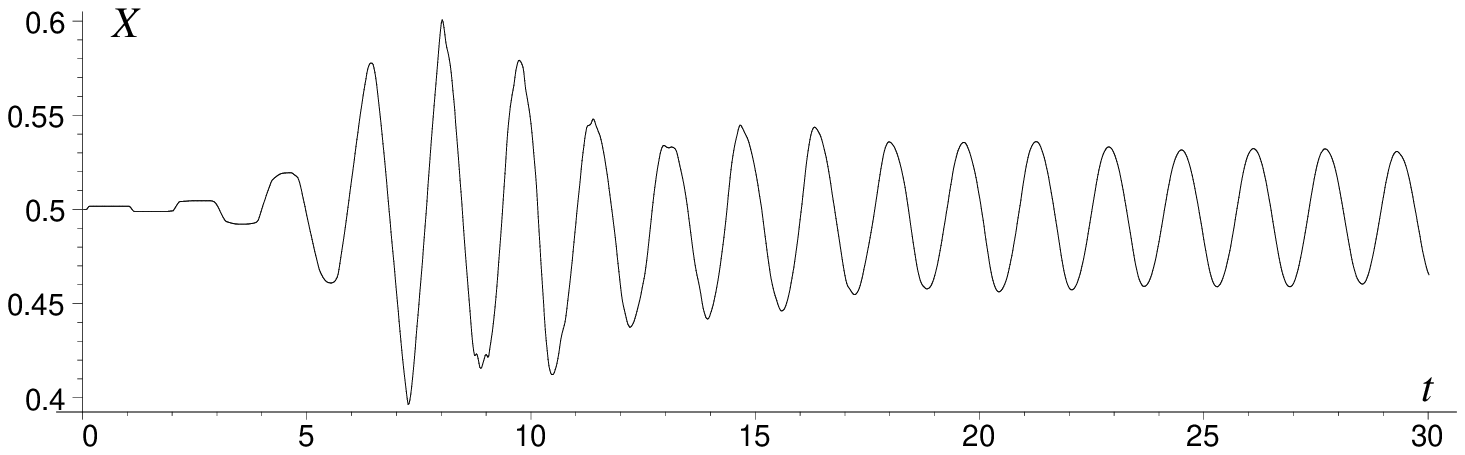}}

\begin{center}
Figure 7: Piston's trajectory from the HE for a perturbed rectangular
density with $r=1/2$.
\end{center}
\vspace*{5mm}

The behavior shown on Fig.~7 persists when various perturbations of the
initial density (\ref{p12}) are applied. It seems like there is a
periodic cycle or an invariant manifold of quasi-periodic solutions of
(H1)--(H4) that acts as an attractor. Of course, due to the
time-reversibility of the hydrodynamical equations there can be no
attractors in the strict sense. It is more likely that there is an
invariant manifold of periodic or quasi-periodic solutions that acts as
a saddle point in the phase space: typical trajectories approach that
manifold temporarily and then slowly move away. We cannot rigorously
prove the existence of periodic or quasi-periodic solutions, but we
construct such solutions by using perturbative analysis.

We will be looking for solution of the hydrodynamical equations
(H1)--(H4) such that the piston makes harmonic oscillations
\be
    X(t)={1\over 2}+ \varepsilon \cos \omega t,
    \ \ \ \ \ \
    \dot{X}(t)= -\varepsilon \omega \sin \omega t,
       \label{XX}
\ee
with some fixed $\omega>0$ and small $\varepsilon>0$. We will
approximate such solutions up to the first order in $\varepsilon$,
i.e.\ ignoring terms of higher order.

The construction is done in two steps. First, we assume that the piston
moves as prescribed by (\ref{XX}) and consider the motion of a fluid
point bouncing against the moving piston $X(t)$ and the fixed wall
$x=0$. Second, we define the density $p(x,v,t)$ which, coupled with the
piston's oscillations (\ref{XX}), satisfies equations (H1)--(H4).

Let the piston move according to Eqs.\ (\ref{XX}). Then fluid points in
the left compartment $0<x<X(t)$ bounce against the wall $x=0$ and the
piston, the latter simply acts on them as a moving wall. It is known
that the phase space of gas particles bouncing against a periodically
moving wall necessarily contains many invariant curves. Moreover, the
region corresponding to high velocities $|v|>v_0$ is densely filled by
such invariant curves, the larger $v_0$ the higher the density of
invariant curves. This fact is a consequence of KAM theory, it was
first proved by R.~Douady in his thesis \cite{Do} and later
independently by S.~Laederich and M.~Levi \cite{LL}. We describe these
invariant curves approximately, up to the first order in $\varepsilon$,
by equation
\be
     v+\varepsilon F(t,V)=V+{\cal O}(\varepsilon^2)
       \label{vF}
\ee
where $v$ denotes the velocity of the particle when it kicks the
piston, $t$ the the collision time, and $V$ is the parameter of the
curve. In fact, we will construct invariant curves for all $V>V_0$ with
some $V_0>0.$ Here we only consider particles to left to the piston,
the particles to the right of the piston are completely symmetric.

Let us consider successive collisions of a gas particle with the
piston. Denote by $v_n>0$ the velocity of the particle before its
$n$-th collision and by $t_n$ the time of that collision. Then the law
of elastic impact reads
\be
   v_{n+1}=v_n -2 \dot{X}(t_n)=
   v_n+2\varepsilon\omega \sin \omega t_n
     \label{E3}
\ee
Let $t_{n+1/2}$ denote the time at which the particle bounces off the
wall $x=0$ between its $n$-th and $(n+1)$-st collisions with the
piston. Obviously, $t_{n+1/2}=t_n+X(t_n)/v_{n+1}$ and
$v_{n+1}(t_{n+1}-t_{n+1/2})=X(t_{n+1}).$ Since we are interested in
knowing $v_n$ up to terms ${\cal O}(\varepsilon),$ it is sufficient to
find $t_n$ up to terms ${\cal O}(1).$ This is easy:
\be
   t_{n+1}=t_n+\frac{1}{v_{n+1}}+{\cal O}(\varepsilon)
   =t_n+{1\over V}+{\cal O}(\varepsilon)
      \label{E4}
\ee
where we used (\ref{vF}).

Now let us look for an invariant curve of the form
$$
     v+\varepsilon F(t,V)= V+{\cal O}(\varepsilon^2)
$$
We have to impose the constraint
\be
   v_{n+1}+\varepsilon F(t_{n+1},V)=
   v_n+\varepsilon F(t_n,V)+{\cal O}(\varepsilon^2)
      \label{E6}
\ee
By equations (\ref{E3}), (\ref{E4}) and (\ref{E6})  we get
$$
    v_n+2\varepsilon\omega \sin\omega t_n+
    \varepsilon F(t_n+1/V+{\cal O}(\varepsilon),V)
    =v_n+\varepsilon F(t_n,V)
$$
Cancelling $v_n$ and $\varepsilon$ and removing the index $n$
gives a general equation for an invariant curve:
\be
   2\omega \sin\omega t+ F(t+1/V,V) - F(t,V) = 0
     \label{E8}
\ee
We construct solutions of this equation in the form
\be
   F(t,V)=a \cos\omega t+b \sin\omega t
     \label{E9}
\ee
where $a$ and $b$ depend on $V$. By substituting this expression into
(\ref{E8}) we find that (\ref{E8}) can only hold if
\begin{eqnarray*}
    a(\cos(\omega/V)-1)+b\sin(\omega/V) &=& 0\\
  a\sin(\omega/V) - b(\cos (\omega/V)-1) &=& 2\omega
\end{eqnarray*}
The solution of the above system is
\begin{eqnarray*}
   a &=& \frac{\omega\sin(\omega/V)}{1-\cos(\omega/V)}\\
   b &=& \omega
\end{eqnarray*}

\medskip\noindent{\em Remark.} Notice that $a$ (and hence the invariant
curve) is not defined for $V=\frac{\omega}{2\pi k}$, $k=\pm 1,\pm
2,\ldots$ To avoid these singularities, we will not use invariant
curves corresponding to $V\leq \frac{\omega}{2\pi}$. In particular, the
density $p(x,v,t)$ that we define below will be constant for $|v|\leq
\frac{\omega}{2\pi}$. \medskip

Thus, for any $V>V_0>{\omega\over 2\pi}$ we can define an
invariant curve $u(x,t;V)$ in the phase space of gas particles,
where $V$ is the parameter of the curve and $u(x,t;V)$ is the
velocity of the particle on the curve at point $x$ at time $t$.
The curve is made by two branches: the upper branch $u^+$ and the
lower branch $u^-.$ Obviously, we have
\begin{eqnarray*}
  u^+(X(t),t;V) &=& V-\varepsilon F(t,V)+{\cal O}(\varepsilon^2)\\
  u^-(X(t),t;V) &=& -[V-\varepsilon F(t,V)-2 \dot{X}(t)]+
  {\cal O}(\varepsilon^2)\\
  u^+(t,0,V) &=& -u^-(t,0,V)
\end{eqnarray*}
Note that the last equation here is equivalent to (\ref{E6}).

Now we define a density $p(x,v,t)$ so that its value on each invariant
curve $u(x,t;V)$, $|V|>V_0$, is a constant denoted by $\rho(V)$.
Between the curves $u^+(x,t;V_0)$ and $u^-(x,t;V_0)$ we set the density
to a constant equal to one. Therefore
$$
  p(x,u^+(t,x,V),t) = p(x,u^-(t,x,V),t) = \rho(V)
   \ \ \ \ {\rm if}\ \ \  V>V_0
$$
and
$$
       p(x,v,t) \equiv 1\ \ \ \ \ \ {\rm if}\ \ \
       u^-(x,t;V_0)<v<u^+(x,t;V_0)
$$
The function $\rho(V)$ and the ``cutoff'' value $V_0>\frac{\omega}
{2\pi}$ will be specified below.

\medskip\noindent{\bf Example.} Let us set $\rho(V)\equiv 0$ for $V>V_0$, i.e.
$$
    p(x,v,t)=\left\{ \begin{array}{ll}
    1 & {\rm for}\ \ u^-(x,t;V_0)<u<u^+(x,t;V_0)\\
    0 & {\rm elsewhere}
    \end{array}\right .
$$
In order to compute the pressure on the piston we only need to
know the density $p(x,v,t)$ at the point $x=X(t)$, i.e.\ we need
to know the function
$$
    v(t,V):=u^+(X(t),t,V)=V+\varepsilon F(t,V)+{\cal O}(\varepsilon^2)
$$
In our example the density on the piston (on the left hand side) is $1$
up to $v^+=V_0-\varepsilon F(V_0,t)$. The density on the piston on the
right hand side is $1$ up to a similar invariant curve, which is phase
shifted by $\Delta t=\pi/\omega$. Therefore the density on the right
hand side is $1$ down to $v^-=-V_0-\varepsilon F(t,V_0).$ Since
$F(t+\pi/\omega)=-F(t)$ by (\ref{E9}), the velocity of the piston is
exactly the average of $v^+$ and $v^-$ and therefore is $-\varepsilon
F(t,V_0)$.

Thus our density and the piston satisfy the hydrodynamical equations
(H1)--(H4) if $\dot{X}=-\varepsilon F(t,V_0)$, which gives
$$
   -\varepsilon\omega\sin\omega t=
   -\varepsilon\omega\sin\omega t
   -\frac{\varepsilon\omega\sin(\omega/V_0)\cos\omega t}
   {1-\cos(\omega/V_0)}
$$
In our example, the only possible choice is $V_0=\frac{\omega}{\pi}.$
\medskip

Now let us consider the case of a generic function $\rho(V).$ The
pressure on the piston on the left hand side is equal to
$$
    P_L=\int_{\dot{X}}^{\infty} p_L(v)\,(v-\dot{X})^2\, dv=
    \int_0^{\infty} p_L(v)\,(v^2-2v\dot{X})\, dv
    +{\cal O}(\varepsilon^2)
$$
where $p_L(v)=p(X(t)-0,v,t)$ is the density on the piston (we have used
the fact that $\dot{X}={\cal O}(\varepsilon)$). Recall that the density
is $p_L(v)=\rho(V)=\rho(v+\varepsilon F(t,v))+{\cal O}(\varepsilon^2).$
>From now on we neglect terms of order ${\cal O}(\varepsilon^2)$. Then
we get
$$
   P_L=\int_0^{\infty}
   (v^2-2v\dot{X})\,\rho(v+\varepsilon F(t,v))\, dv
$$
The pressure on the right hand side is given, by analogy,
$$
   P_R=\int_{-\infty}^0
   p_R(v)\,(v^2-2v\dot{X})\, dv=
   \int_0^\infty p_R(-v)\,(v^2+2 v \dot{X})\, dv
$$
Note that for $v>0$ we have $p_R(v)=\rho(V)=\rho(v+\varepsilon
F(v,t+\pi))=\rho(v-\varepsilon F(t,v))$. Therefore
$$
    P_R=\int_0^\infty (v^2+2v\dot{X})\,\rho(v-\varepsilon F(t,v))\, dv
$$
We now conclude that $P_L=P_R$ iff
$$
    \dot{X}=\varepsilon\,\frac{\int_0^{\infty}\rho'(v)F(t,v)v^2\, dv}
     {\int_0^\infty \rho(v)\, 2 v\, dv}=
     -\varepsilon\, \frac{\int_0^{\infty}\rho'(v)F(t,v)v^2\, dv}
     {\int_0^\infty \rho'(v)v^2\, dv}
$$
which is analogous to our early formula (\ref{Vt}).

Using (\ref{E9}) and the subsequent equations we find
$$
    \dot{X}= -\varepsilon \omega \sin\omega t-
    \varepsilon\,\frac{\int_0^\infty \rho'(v)
    \,\frac{\sin(\omega/v)}{1-\cos(\omega/v)}\,v^2\, dv}
    {\int_0^\infty \rho'(v)\,v^2\, dv}\,
    \omega \cos\omega t
$$
Our density, coupled with the piston oscillations (\ref{XX}), satisfies
the hydrodynamical equations (H1)--(H4) if and only if
$\dot{X}=-\varepsilon \omega \sin\omega t$. This implies
\be
   \int_{\omega/2\pi}^\infty dv\,\rho'(v)\,v^2\,
   \frac{\sin(\omega/v)}{1-\cos(\omega/v)}\,=0
      \label{E24}
\ee
where we have imposed $\rho'=0$ for $v\leq\omega/2\pi.$

Interestingly, (\ref{E24}) seems to be related to our early equation
(\ref{zexp}). Precisely, let $z$ in (\ref{zexp}) be a purely imaginary
number, $z=\omega i$. Also note that $\rho(v)$ in (\ref{zexp}) is just
proportional to $v^2\rho'(v)$ here. Then (\ref{E24}) becomes equivalent
to Im$\, F(z)=0$, with $F(z)$ defined by (\ref{F(z)}). In other words,
(\ref{E24}) expresses the ``imaginary part'' of the equation
(\ref{zexp}). We already observed in the previous section that Im$\, z$
characterized the frequency of oscillations of unstable perturbations,
and here $\omega=\,$Im$\, z$ is the frequency of oscillations of the
piston. We note that for $z=\omega i$ one always has Re$\, F(z)=0$, as
it follows from (\ref{F(z)}), hence in our case (\ref{E24}) is
equivalent to $F(z) =0$.

Next, we note that  $\rho'(v)$ must be supported on both the intervals
$(\omega/2\pi, \omega/\pi]$ and $[\omega/\pi, +\infty)$. In fact the
fraction in Eq.\ (\ref{E24}) is negative for $v\in (\omega/2\pi,
\omega/\pi)$ and positive for $v>\omega/\pi.$

Obviously Eq.\ (\ref{E24}) can be satisfied in many different ways, all
of them leading to different solutions of the hydrodynamical equations.

Lastly, we want to examine a special case when the initial density
$p(x,v,0)= \rho(v)$ is a monotonic function in $|v|$, i.e.\ when the
hydrodynamic equations are stable. Then there are some quantitative
restrictions on the period of oscillations of the piston. The period
$T=2\pi/\omega$ can be bounded from below by a function of the average
kinetic energy $\la K\ra=K/M,$ where
$$
   K=\int_0^\infty \rho(v) {v^2\over 2}\, dv
   \ \ \ \ \ \ {\rm and}\ \ \ \ \ \
   M=\int_0^\infty \rho(v)\, dv
$$

\begin{proposition}
Let $\rho'\leq 0$ be supported on the interval $[\omega/2\pi,\infty)$
and satisfy Eq.\ (\ref{E24}). Then the period of oscillations $T$ is
bounded by
$$
      T\geq \sqrt{\frac{2}{3\la K\ra}}
$$
The equality holds when $\rho'= (\pi/\omega) \delta(v-\omega/\pi)$,
i.e.\ when $\rho(|v|)$ is constant on $(0,\omega/\pi)$ and $0$
elsewhere. \label{prE}
\end{proposition}

Even though the proposition is stated for monotonic densities only, let
us apply it to our unstable density (\ref{p12}). In this case $K=7/24$
and $T=\sqrt{16\over 7}\simeq 1.51186$. The experimentally determined
period of oscillations of the piston is $T\simeq 1.62$, see \cite{CL}.
We also simulated the piston trajectory with other unstable densities
(\ref{p1r}) with $r\to 0$ and observed that the period of oscillations
approached its lower bound 2 given by the above proposition.

\medskip\noindent{\em Proof of Proposition~\ref{prE}.} Consider a
function
$$
   G(v):=v\,\frac{\sin(\omega/v)}{1-\cos(\omega/v)}
$$
Then equation (\ref{E24}) reads
$$
    C:=\int_{\omega/2\pi}^\infty - \rho' v  G \, dv=0
$$
Note that in the interval $[\omega/2\pi,\infty)$ the function $G(v)$
is strictly increasing and that $G(\omega/2\pi)=-\infty$,
$G(\omega/\pi)=0$, and $G(\infty)=\infty$.

Introducing a new function $R(v)=-\rho'(v)v\geq 0$ and integrating by
parts yields
$$
   M=\int_{\omega/2\pi}^\infty R(v)\,  dv,\ \ \ \
   K=\int_{\omega/2\pi}^\infty R(v)\frac{v^2}{6}\, dv,\ \ \ \
   C=\int_{\omega/2\pi}^\infty R(v) G(v)\, dv
$$

It is useful to replace $v$ by a new variable $u=G(v)$,
$-\infty<u<\infty$. Since $G$ is strictly increasing, we can write
$$
    M=\int_{-\infty}^\infty S(u)\, du,\ \ \ \
    K=\int_{-\infty}^\infty S(u) \eta(u)\, du,\ \ \ \
    C=\int_{\infty}^\infty S(u) u\, du
$$
where $S(u)=R(G^{-1}(u)) / G'(G^{-1}(u))$ and $\eta(u)= (G^{-1}(u))^2/
6$ (here $G^{-1}$ denotes the inverse of the function $G$).

We have to solve the following variational problem: minimize $K/M$
under the constraint $C=0.$ As we shall see in the sequel the function
$\eta$ turns out to be convex. This easily implies
Proposition~\ref{prE}.

Also, the convexity of $\eta$ implies that the solution of the
variational problem $\bar{S}$ is a delta-function centered at $u=0,$
i.e.\ at $v=\omega/\pi$ (so that $C$ vanished).

So it only remains to prove that $\eta$ is a convex function of $u.$ By
direct computation we get
$$
   6\eta'=\frac{d}{du}(G^{-1}(u))^2=2
   G^{-1}(u)\frac{d G^{-1}(u)}{du}=2v/G'(v)
$$

Hence it is sufficient to prove that the function $G'(v)/v$ is strictly
decreasing in the interval $v>\omega/2\pi.$ Without loss of generality
we set $\omega=1$, then
$$
    G(v)={v \sin(1/v)\over 1-\cos(1/v)}
$$
Consider a new function
$$
    H(v):=\frac{G'(v)}{v}={1+v \sin(1/v)\over v^2(1-\cos(1/v))}
$$
then
$$
    H(v)'=\frac{-v -(-1+v^2)\sin(1/v)+v \cos(1/v)(1+v \sin(1/v))}
    {v^4(1-\cos(1/v))^2}
$$
The denominator of $H'$ being positive, we only need to prove that the
numerator of $H'$ is negative in the interval $v>1/2\pi.$

If we replace $v$ by $1/x$ and multiply by the numerator by $x^2$ we
find the expression
\begin{eqnarray*}
    h(x) &=& -x+(x^2-1)\sin(x)+\cos(x)(x+\sin x)\\
     &=& (\cos x-1)(\sin x + x)+x^2 \sin x
\end{eqnarray*} We need to show that $h(x)<0$ in the interval
$x\in(0,2\pi).$ First of all, $h(0)=h(2\pi)=0$ and for any
$x\in(\pi,2\pi)$ the expression is clearly negative.

It only remains to prove that $h(x)$ is negative in $(0,\pi].$ By
computing  the Taylor expansion of $h$ about $h=0$ one finds
$$
   h(x)=\sum_{k=3}^{+\infty}(-1)^k\,\frac{2^{2k}-4 k^2}
   {(2k+1)!}\, x^{2 k+1}
$$
It is easy to prove that for any $x\in(0,\pi]$ this is an alternating
series, the absolute values of its terms being strictly decreasing.

The first few terms of the above expansion are
$$
   h(x)=-{x^7\over 180}\left(1-{2\over 21}x^2+{1\over 240}x^4-
    {19\over 166320}x^6 \right)+{\cal O}(x^{15})
$$
Therefore
$$
    -{x^7\over 180}< h(x)<-{x^7\over 180}
    \left(1-{2\over 21} x^2\right)
$$
which implies that $h(x)<0$ for any $x\in(0,\pi].$ $\Box$

\medskip\noindent
{\bf Acknowledgements}. We thank E. Presutti and Ya. Sinai for their
contributions to the early stages of this work, C. Marchioro, M.
Pulvirenti, and N. Simanyi for useful discussions. We also thank Weinan
E, Rick Falk and Michael Vogelius for helpful suggestions about the
numerical solution of the hydrodynamical equations. E. Caglioti was
partially supported by MIUR GNFM and by the European network HYKE,
contract HPRN-CT-2002-00282. N.~Chernov was partially supported by NSF
grant DMS-0098788. J.~Lebowitz was partially supported by NSF grant
DMR-9813268 and by Air Force grant F49620-01-0154. This work was
started when the authors stayed at the Institute for Advanced Study
with partial support by NSF grant DMS-9729992.

\end{document}